\documentclass[preprint,amssymb,amsmath,amssymb,amsfonts,superscriptaddress,showpacs]{revtex4-1}

\usepackage{graphicx,graphics,pstricks,epsfig,wrapfig}

\begin{document}
\DeclareGraphicsExtensions{.pdf,.eps,.epsi,.jpg}

\title{Updated Summation Model: An Improved Agreement with the Daya Bay Antineutrino Fluxes}

\author{M. Estienne}
\affiliation{SUBATECH, CNRS/IN2P3, Universit\'e de Nantes, Institut Mines-Telecom de Nantes, F-44307 Nantes, France}

\author{M. Fallot}
\affiliation{SUBATECH, CNRS/IN2P3, Universit\'e de Nantes, Institut Mines-Telecom de Nantes, F-44307 Nantes, France}

\author{A. Algora}
\affiliation{IFIC (CSIC-Univ. Valencia), Valencia, Spain}
\affiliation{Institute of Nuclear Research, Debrecen, Hungary}

\author{J. Briz-Monago}
\affiliation{SUBATECH, CNRS/IN2P3, Universit\'e de Nantes, Institut Mines-Telecom de Nantes, F-44307 Nantes, France}

\author{V.M. Bui}
\affiliation{SUBATECH, CNRS/IN2P3, Universit\'e de Nantes, Institut Mines-Telecom de Nantes, F-44307 Nantes, France}

\author{S. Cormon}
\affiliation{SUBATECH, CNRS/IN2P3, Universit\'e de Nantes, Institut Mines-Telecom de Nantes, F-44307 Nantes, France}

\author{W. Gelletly}
\affiliation{Department of Physics, University of Surrey, GU2 7XH, Guildford, UK}

\author{L. Giot}
\affiliation{SUBATECH, CNRS/IN2P3, Universit\'e de Nantes, Institut Mines-Telecom de Nantes, F-44307 Nantes, France}

\author{V. Guadilla}
\affiliation{SUBATECH, CNRS/IN2P3, Universit\'e de Nantes, Institut Mines-Telecom de Nantes, F-44307 Nantes, France}

\author{D. Jordan}
\affiliation{IFIC (CSIC-Univ. Valencia), Valencia, Spain}

\author{L. Le Meur}
\affiliation{SUBATECH, CNRS/IN2P3, Universit\'e de Nantes, Institut Mines-Telecom de Nantes, F-44307 Nantes, France}

\author{A. Porta}
\affiliation{SUBATECH, CNRS/IN2P3, Universit\'e de Nantes, Institut Mines-Telecom de Nantes, F-44307 Nantes, France}

\author{ S. Rice}
\affiliation{Department of Physics, University of Surrey, GU2 7XH, Guildford, UK}
\author{ B. Rubio}
\affiliation{IFIC (CSIC-Univ. Valencia), Valencia, Spain}

\author{J. L. Ta\'{\i}n}
\affiliation{IFIC (CSIC-Univ. Valencia), Valencia, Spain}

\author{ E. Valencia}
\affiliation{IFIC (CSIC-Univ. Valencia), Valencia, Spain}

\author{A.-A. Zakari-Issoufou}
\affiliation{SUBATECH, CNRS/IN2P3, Universit\'e de Nantes, Institut Mines-Telecom de Nantes, F-44307 Nantes, France}

\begin{abstract}

A new summation method model of the reactor antineutrino energy spectrum is presented. It is updated with the most recent evaluated decay databases and with our Total Absorption Gamma-ray Spectroscopy measurements performed during the last decade. For the first time the spectral measurements from the Daya Bay experiment are compared with the detected antineutrino energy spectrum computed with the updated summation method without any renormalisation. The results exhibit a better agreement than is obtained with the Huber-Mueller model in the 2 to 5 MeV range, the region which dominates the detected flux. An unexpected systematic trend is found that the detected antineutrino flux computed with the summation model decreases with the inclusion of more Pandemonium free data. The detected flux obtained now lies only 1.9\% above that detected in the Daya Bay experiment, a value that may be reduced with forthcoming new Pandemonium free data leaving less and less room to the reactor anomaly. 
Eventually, the new predictions of individual antineutrino spectra for the $^{235}$U, $^{239}$Pu, $^{241}$Pu and $^{238}$U are used to compute the dependence of the reactor antineutrino spectral shape on the fission fractions.

\end{abstract}
\maketitle

Neutrino oscillation parameters have been precisely measured in recent decades confirming a three-flavour scheme and paving the way for future measurements of the CP-violation phase, the mass hierarchy and an understanding of the nature of the neutrino particle~\cite{Juno, DoubleBeta1, DoubleBeta2, DoubleBeta3, DoubleBeta4}. 
Nevertheless, some neutrino experiments have obtained results that cannot be explained by the three-flavour oscillation model~\cite{LSND, MiniBoone, Mention, Gallium}. 
One of these anomalies, the "reactor anomaly" denotes the deficit observed between the detected antineutrino flux in reactor neutrino experiments at less than 100~m from reactors with 
respect to new predictions of this detected flux~\cite{Mention}. 
These predictions have been obtained by an improved conversion~\cite{Mueller, Huber} of the integral beta energy spectra measured by Schreckenbach {\it et al.}~\cite{SchreckU5-1, SchreckU5-2, SchreckU5Pu9,Hahn}. It led to a new normalisation of the detected antineutrino flux lying some 6\% above the detected flux at short distances from reactors~\cite{Mention}, 
triggering the search for sterile neutrinos close to experimental reactors~\cite{WhitePaper}. Meanwhile, the reactor experiments Double Chooz~\cite{DC}, Daya Bay (DB)~\cite{DayaBay} and Reno~\cite{RENO} have measured the shape of the reactor antineutrino spectrum close to Pressurized Water Reactors (PWR). The comparison between the converted spectra and the measured spectra not only confirmed the reactor anomaly, but also exhibited a large distortion of the data with respect to the model between 5 and 7 MeV in antineutrino energy. 
These two findings raised questions about the antineutrino predictions based on the conversion method, hereinafter called the Huber-Mueller model (H-M)~\cite{Huber, Mueller} and  the associated systematic uncertainties of several nuclear effects~\cite{Huber,Hayes,FangBrown,Hayen}.

In 2017, the Daya Bay experiment published the first measurement of the evolution of the antineutrino flux with fuel burnup~\cite{DayaBay2017}. They compared it with the changing flux obtained with the H-M model. Overall the slope of the evolution of the flux with an increasing percentage of fissions from $^{239}$Pu is rather well reproduced by the model, although the absolute magnitude of the computed flux is higher than observed by 6\% reinforcing the reactor anomaly. Furthermore, the Daya Bay collaboration could disentangle the contributions to the antineutrino flux from $^{235}$U and $^{239}$Pu fission. In comparison with the predictions from the H-M model, good agreement was found with the flux arising from the fission of $^{239}$Pu, while the disagreement in flux arising from the fission of $^{235}$U could almost explain the experimental deficit by itself. Although  statistics are needed, 
this result would indicate that the reactor anomaly could not be explained by any neutrino oscillation since this should be independent of the fuel and that there is a potential problem in the $^{235}$U antineutrino spectrum.

An alternative to the H-M model is the summation method (SM). It consists of summing all the individual beta branches composing the total antineutrino or beta spectrum weighted by the beta decay activities. The SM model is the only tool which allows one to study the components of the reactor antineutrino energy spectrum and to predict reactor antineutrino spectra over the full energy range from any fuel and under any irradiation conditions. This method, originally proposed in~\cite{King} followed by~\cite{Avignonne} and then by~\cite{Vogel81,Tengblad} relies completely on the available nuclear data of the fission yields combined with the beta decay data for the fission products. It was revisited at the same time as the conversion method~\cite{Mueller}, leading to the conclusion that the Pandemonium effect~\cite{Hardy77} affects some of the nuclear beta decay data that play an important role in the calculation of the antineutrino energy spectrum~\cite{Fallot}. 
The Pandemonium effect is a systematic uncertainty encountered in beta decay studies using Germanium detectors. 
Because of the limited efficiency of these detectors gamma-rays de-exciting high energy levels may not be detected and hence the beta branching to these levels is underestimated. 
The result is an overestimate of the high energy part of the antineutrino energy spectra. 
The use of the Total Absorption Gamma-ray Spectroscopy (TAGS) technique~\cite{RubioGelletly} allows one  
to correct for these systematic uncertainties. 
This effect is the major bias in the determination of the antineutrino spectra with the SM, well above the effect of forbidden non-unique transitions~\cite{Hayes,FangBrown,Hayen}. The correction of most of the data affected by Pandemonium is thus an essential pre-requisite for the calculation of its associated uncertainties. 
{Since~\cite{Mueller}}, we have performed TAGS experimental campaigns leading to the correction of fifteen nuclear decays of major relevance for the reactor antineutrino spectra.
Though their relative impacts have been evaluated in~\cite{Fallot,PRLAlgora,Zak,Valencia,Rice,Guadilla2018}, the resulting absolute detected antineutrino energy spectrum after their inclusion has never been studied.
In this paper, we have assessed for the first time this spectrum  without any renormalisation computed with an updated SM model. 
Our purpose is to: (1) compare it with the measurements from Daya Bay and with the H-M model, (2) show the unexpected systematic impact of including more Pandemonium free data in the SM on the detected flux, (3) quantify the new discrepancy between the Daya Bay measurements of the antineutrino flux and that obtained with the model in a fuel dependent way and (4) provide the community with this new SM model since it shows improved agreement with the neutrino measurements.

The main characteristics of our model presented in~\cite{Fallot} have been updated in two essential aspects. Firstly the cocktail of the beta decay data used in the calculation has been updated and taken in 
the following order of priority after the TAGS data and the data from~\cite{Tengblad}: JEFF-3.3~\cite{JEFF}, ENDF/B-VIII.0~\cite{ENDF}, the gross theory from~\cite{GrossTheo2018} and for the remaining nuclei the Qbeta approximation presented in~\cite{Fallot}. Secondly several SM models are defined depending on the included TAGS nuclear decays in the following way.
The SM model taking into account TAGS results for $^{102,104,105,106,107}$Tc, $^{105}$Mo and $^{101}$Nb~\cite{PRLAlgora} will be called SM-2012~\cite{Fallot} hereinafter. The additional inclusion of $^{92,94}$Rb and $^{87,88}$Br~\cite{Zak,Valencia} and of $^{91}$Rb, $^{86}$Br~\cite{Rice} will lead to the SM-2015 and SM-2017 models respectively. Eventually, the SM-2018 model takes moreover into account TAGS data for $^{100, 100m, 102, 102m}$Nb~\cite{Guadilla2018} that 
were found to have a large impact on the antineutrino spectra. 
In our calculations, the weak magnetism correction is taken into account after~\cite{Huber} except for the 0$^-$ to 0$^+$ transitions. All the energy spectra presented in this article correspond to antineutrino energy.

\begin{figure}[ht]
  \centering
\includegraphics[width=0.70\textwidth]{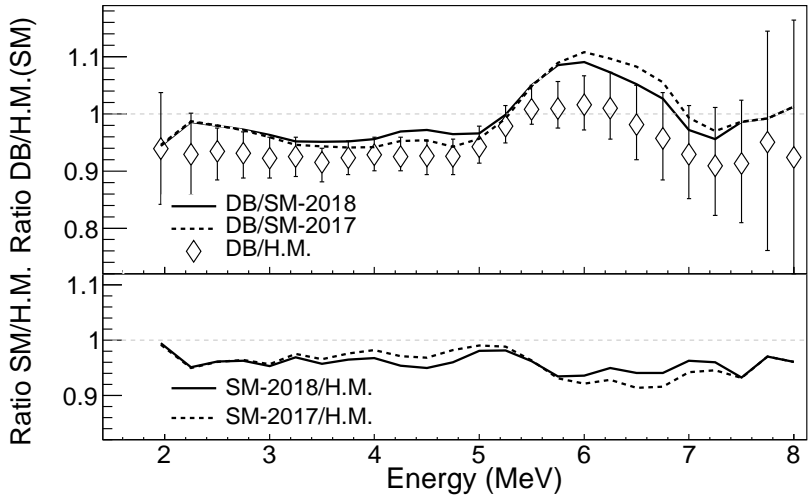}
 \caption{Upper Panel: Ratio of the DB antineutrino energy spectrum to that of the H-M model as in~\cite{An2017} (open diamonds), to that of the SM-2017 (dashed line) and the SM-2018 (continuous line) models (see text). Lower Panel: Ratio of the SM-2017 (dashed line) and SM-2018 (continuous line) antineutrino energy spectra to that of the H-M model~\cite{Mueller, Huber}.}
  \label{Fig:Figure2}
\end{figure}
In the top panel of Fig.~\ref{Fig:Figure2}, we show the ratio of the detected antineutrino energy spectrum published by Daya Bay to the antineutrino spectrum computed with the H-M model, using the fission fractions given in~\cite{An2017} (open diamonds). The antineutrinos are detected through the Inverse Beta Decay process (IBD) and the cross-section used to compute the detected antineutrinos is the one from~\cite{VogelBeacom}. 
The ratios of the Daya Bay spectrum over the spectra predicted with the SM-2017 (dashed line) and SM-2018 (continuous line) models are also displayed~\cite{extraMat}. In these calculations, the fission yields have been computed after 450~days of irradiation which represents roughly the average of the ages of the assemblies in the core of a standard PWR.
The global shape of the DB over SM ratios are similar to that of DB over H-M, but closer to one except in the 5 to 7~MeV range. The inclusion of the TAGS data for the niobium isotopes improves the situation above 3~MeV, extending the good agreement in shape with the Daya Bay spectrum up to 5~MeV. This is worth noting since the 2 to 5 MeV energy region dominates the detected antineutrino flux. There is still a shape difference between the Daya Bay spectrum and the SM-2018 antineutrino spectrum between 5 and 7~MeV, but its amplitude is reduced by the inclusion of the new data which improve the agreement globally w.r.t. SM-2017. 
In the lower panel of the figure, the ratios of the detected summation method spectra SM-2017 (dashed line) and SM-2018 
(plain line) to the H-M model spectra are displayed for comparison. 
In the case of the 2 to 3 MeV energy region, the agreement with H-M is equally good for the 2017 or 2018 versions of the SM. Above 3~MeV, the ratio of the two models is rather flat and normalized on average about 3-4\% below one.
Overall the SM-2018 model shows a globally improved agreement with the shape of the converted spectrum but not with its normalisation. An important conclusion is that the reactor antineutrino energy spectrum obtained with the SM exhibits a normalisation in the 2 to 5 MeV range  more compatible with the Daya Bay results than the H-M model, but it does not explain the shape anomaly in the 5 to 7 MeV region.
\begin{figure}[ht]
  \centering
\includegraphics[width=0.7\textwidth]{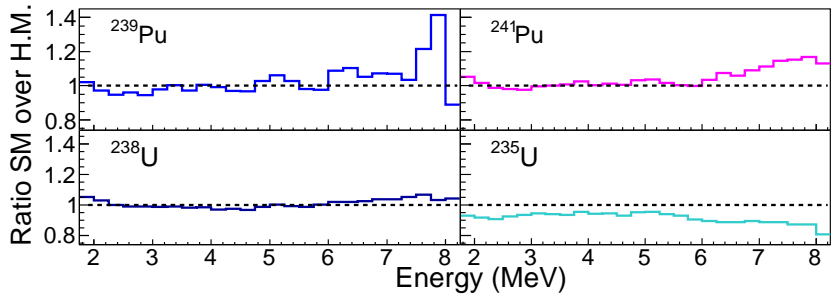}
\caption{Ratios of the antineutrino energy spectra obtained with the SM-2018 model with the converted spectra for $^{239}$Pu, $^{241}$Pu, $^{235}$U~\cite{Huber} and Mueller's prediction for $^{238}$U~\cite{Mueller}.}
  \label{Fig:Figure3}
\end{figure}

In Fig.~\ref{Fig:Figure3} the comparisons between the individual antineutrino spectra of the SM-2018 and those of the H-M model for the four main contributions to the total number of fissions in a PWR are shown. The SM spectra are taken at a time corresponding to the irradiation times of Schreckenbach {\it et al.}'s experiments 
for $^{235}$U,  $^{239}$Pu, and $^{241}$Pu 
while the SM spectrum for  $^{238}$U is computed after 450 days of irradiation as was the case for Mueller's spectrum~\cite{Mueller}. 
The normalisation of the SM spectra with respect to the H-M ones seem to confirm the Daya Bay result of~\cite{DayaBay2017} in which the antineutrino deficit is mainly produced by the $^{235}$U contribution to the antineutrino flux. The individual spectra show an improved shape agreement with the Huber spectra exhibiting ratios that are flatter between 2 and 6 MeV
and closer to one for both plutonium isotopes than in~\cite{Fallot}.

The Daya Bay experiment has also measured the detected reactor antineutrino flux normalized per fission as a function of the percentage of fissions of $^{239}$Pu ($F_{239}$).
It is defined as the detected Inverse Beta Decay (IBD) yield in~\cite{DayaBay2017} following the equation:
\begin{equation}
\sigma_f (F_{239}) = \bar{\sigma}_f + \frac{d\sigma_f}{dF_{239}} (F_{239} - \bar{F}_{239}) \\
\end{equation}
The average IBD yield $\bar{\sigma}_f$ is obtained by folding the IBD cross-section with the total antineutrino energy spectrum computed by weighting the $^{235}$U, $^{238}$U, $^{239}$Pu and $^{241}$Pu spectra by their average fission fractions provided in~\cite{DayaBay2017}. 
$\bar{F}_{239}$ is the average $^{239}$Pu fission fraction, and $\frac{d\sigma_f}{dF_{239}}$ is the change of the IBD yield per unit $^{239}$Pu fission fraction. 
Hayes {\it et al.} compared for the first time the IBD results from DB and the H-M model with their own SM calculations based on ingredients which differ from the ones presented here in the amount of the TAGS data included among other things~\cite{Hayes2017}. The authors found a 3.5\% excess between their predictions and DB, {a discrepancy interpreted as a confirmation of the reactor anomaly}.

\begin{figure}[ht]
  \centering
\includegraphics[width=0.65\textwidth]{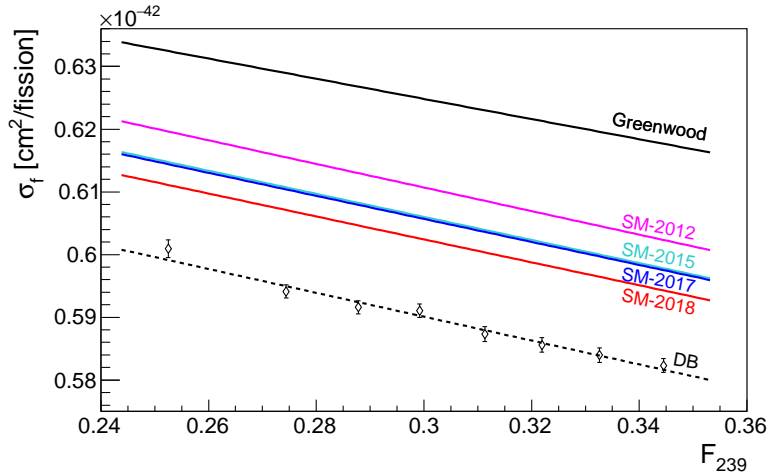}
\caption{IBD yields as a function of the $^{239}$Pu fission fractions for different versions of the summation model (see text). Data (open diamonds) are extracted from~\cite{DayaBay2017}.}
  \label{Fig:Figure5}
\end{figure}

We have reinvestigated the calculations with our prescriptions varying the different TAGS input datasets.
The associated IBD yields as a function of the $^{239}$Pu fission fraction have been computed. They are displayed in Fig.~\ref{Fig:Figure5} and compared with those published by Daya Bay (open diamonds)~\cite{DayaBay2017}. 
The line corresponding to the SM-2012 model is located 3.4\% above the Daya Bay measurements, a value very close to that of~\cite{Hayes2017}.
The IBD yields obtained with the SM-2015 and SM-2017 models have been drawn with the lines labelled with the corresponding names in Fig.~\ref{Fig:Figure5}. One can see that the effect is indeed to decrease the IBD yield, resulting in a difference with the Daya Bay yield of only 2.5\%, quasi-constant over the $F_{239}$ range.
With the SM-2018 model (line correspondingly) 
the remaining difference with the Daya Bay IBD yields 
is reduced to only 1.9\%. 
 These results show that the IBD yield is strongly affected by the use of Pandemonium free data. 
In addition, they show an unexpected systematic trend to decrease the IBD yield when including more Pandemonium free data. 
This observation is supported by the IBD yield obtained by limiting the TAGS data only to those from Greenwood {\it et al.}~\cite{Greenwood} (black line), located well above the others in Fig.~\ref{Fig:Figure5}, which reflects the level of detected flux that would have been obtained using the SM from~\cite{Mueller}.
This important result highlights that the relative discrepancy with the Daya Bay measurements will be reduced even more with forthcoming TAGS results. This flux reduction is the consequence of the distortion of the antineutrino energy spectrum induced by the correction of the Pandemonium effect. 
In order to test the robustness of the IBD yield obtained with the SM, the range of variation of the predicted flux depending on different choices of decay data has been evaluated. 
Some of the Pandemonium free data which exhibit large uncertainties have been replaced by another dataset taken from evaluated databases, most of the time apparently affected by the Pandemonium effect. This operation resulted in a 1.2\% shift upward with respect to the IBD yield found with SM-2018, with a maximum impact of 2~\% from 2~MeV to 3.5~MeV, which puts a stringent constraint on the global normalisation of the SM spectrum. Choosing the ENDF/B-VIII database over the JEFF3.3 one resulted in a 0.6\% shift downward with respect to SM-2018 because of discrepancies in the evaluations. 

\begin{table*}[!t]
\begin{center}
\begin{tabular}{c|c|c|c|c|c|c}
\hline
\hline
\multicolumn{1}{c|}{\textbf {}}  & \multicolumn{1}{c|}{\textbf {DB}}   &\multicolumn{1}{c|}{\textbf {SM 2018}} &\multicolumn{1}{c|}{\textbf {SM 2017}} & \multicolumn{1}{c|}{\textbf {SM 2012}}& \multicolumn{1}{c|}{\textbf {SM from \cite{Hayes2017}}}& \multicolumn{1}{c}{\textbf {H-M}}\tabularnewline \hline
$\sigma_f$ (10-43cm$^2$)&5.9$\pm$0.13&6.01&6.05& 6.10& 6.11&6.22$\pm$0.14\tabularnewline
$\frac{d\sigma_f}{dF_{239}}$(10-43cm$^2$)&-1.86$\pm$0.18 &-1.82&-1.83 &-1.87&-2.05 &-2.46$\pm$0.06 \tabularnewline
$\sigma_5$ (10-43cm$^2$) &6.17$\pm$ 0.17&6.28&6.31& 6.38& 6.49 &6.69$\pm$0.15\tabularnewline
$\sigma_9$ (10-43cm$^2$) &4.27$\pm$ 0.26 &4.42&4.44& 4.47&4.49& 4.36$\pm$0.11\tabularnewline
$\sigma_8$ (10-43cm$^2$) &10.1$\pm$1.0&10.14&10.20& 10.27& 10.2& 10.1$\pm$1.0\tabularnewline
$\sigma_4$ (10-43cm$^2$) &6.04$\pm$0.6&6.23&6.27& 6.29& 6.4& 6.04$\pm$0.6\tabularnewline
$\sigma_5/\sigma_9$ &1.445$\pm$0.097&1.421&1.421& 1.427& 1.445 &1.53$\pm$ 0.05\tabularnewline
\hline
\hline 
\end{tabular}
\end{center}
\caption{The IBD average yields, the variation with the $^{239}$Pu content of the fuel, and the contributions from individual actinides measured by the Daya Bay collaboration (second column), computed with SM 2018, 2017 and 2012 respectively (third, fourth and fifth columns) or with the SM from~\cite{Hayes2017} (sixth column) and with the H-M model (seventh column). The labels 5, 9, 8 and 4 stand for $^{235}$U, $^{239}$Pu, $^{238}$U and $^{241}$Pu respectively.}
\label{table1}
\end{table*}

In table~\ref{table1} the slopes $\frac{d\sigma_f}{dF_{239}}(F_{239})$ as well as the values of $\bar{\sigma}_f$ associated with our model in its versions from 2012, 2017 and 2018 are presented and those taken from the Daya Bay paper along with those from Hayes {\it et al.} and the H-M model.  
The slope associated with the line labelled Greenwood in Fig.~\ref{Fig:Figure5} is $\frac{d\sigma_f}{dF_{239}} = -1.60$ which shows that after the inclusion of the TAGS data in 2012~\cite{Fallot} in our summation calculation the other slopes are almost unaffected by the additional input of TAGS data. They remain close to the experimental value, well into the measured uncertainty. 
The four IBD yields corresponding to the individual contributions to the fissions of $^{235}$U, $^{238}$U, $^{239}$Pu and $^{241}$Pu are also displayed in table~\ref{table1}. 
One can see that the more Pandemonium free results are included, the closer to the Daya Bay results the model gets. 
The situation is in contrast with the IBD yields provided by the H-M model, in which $\sigma_5$ carries most of the flux discrepancy. Overall, the SM-2018 model best reproduces the Daya Bay IBD average yields and slopes, and the $^{235}$U individual contribution simultaneously. The other contributions remain well within the experimental uncertainties and represent the closest to the DB central value ever obtained with a SM model. 

\begin{figure}[ht]
  \centering
 \includegraphics[width=0.7\textwidth]{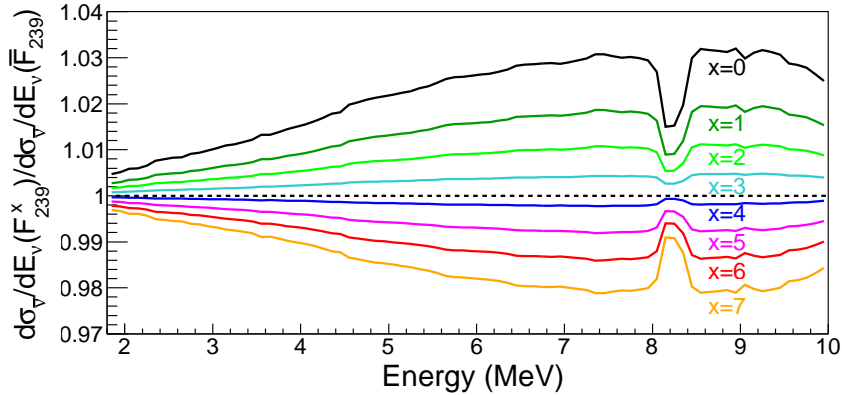}
\caption{Ratios between SM-2018 spectra built in varying the fission fractions (labelled $F^x_{239}$, with x increasing with the $F_{239}$ values ranging from 24.2\% to 34.4\% as given in~\cite{DayaBay2017}) over that built with the average fission fractions.}
  \label{Fig:Figure7}
\end{figure}

In Fig.~\ref{Fig:Figure7}, the ratios between the SM-2018 detected antineutrino spectrum built with the fission fractions 
taken in~\cite{DayaBay2017} (and used in Fig.~\ref{Fig:Figure5}) and that built with the average fission fractions are displayed as a function of antineutrino energy. These ratios show the evolution of the shape of the energy spectrum with the burn-up, that will be measured by the reactor antineutrino experiments when they have accumulated enough statistics. As expected, the ratio of the spectra decreases globally up to 10~MeV with the increasing number of fissions coming from $^{239}$Pu since its antineutrino energy spectrum is lower than that of $^{235}$U. 
Fig.~\ref{Fig:Figure7} also shows that comparing antineutrino energy spectra with different sets of fission fractions, the ratios obtained reveal a positive (negative) slope in the 2 to 8~MeV energy range when the $^{239}$Pu fission fraction is smaller (larger) than the average value taken as a reference.

In conclusion, we have presented an update of our SM with the most recent evaluated decay databases and the inclusion of the TAGS results obtained by our collaboration during the last decade. 
For the first time, the reactor energy antineutrino spectrum predicted by the SM has been compared with that measured by the Daya Bay experiment without any renormalisation. The ratio of the Daya Bay over the updated SM antineutrino energy spectra is improved with respect to the Daya Bay over H-M ratio in the 2 to 5 MeV range. This energy region which dominates the detected flux is the most robust part of the SM energy spectrum with respect to variations in the choice of nuclear data. In view of these results, the predictive power of the SM has been dramatically improved by the inclusion of new TAGS decay data 
paving the way for the computation of the uncertainties associated with the decay data.
The shape distortion between 5 and 7 MeV is still visible and unexplained by the SM to date, but in this energy region the potential impact of remaining Pandemonium affected nuclei or unknown data remains important.
Furthermore, the calculation of the detected antineutrino flux reveals a systematic trend that the reactor anomaly is reduced depending on the Pandemonium free data of the major contributors in a fuel dependent way. The remaining discrepancy with the Daya Bay flux reduces to only 1.9\% and we expect further improvement when more Pandemonium free results are included.
The agreement of the contributions of $^{235}$U, $^{239}$Pu, $^{241}$Pu and $^{238}$U to the detected antineutrino flux and of the slope $\frac{d\sigma_f}{dF_{239}} (F_{239})$ with those of Daya Bay is also improved with our model.
Eventually, a prediction is provided for the absolute detected antineutrino spectrum as a function of the $^{239}$Pu fission fractions for comparison with future experimental results.
The comparison of the updated SM model with the forthcoming measurements of the pure $^{235}$U antineutrino spectra from 
Prospect~\cite{Prospect}, SoLid~\cite{SoLid}, or STEREO~\cite{STEREO} will also provide new insight into the reactor and shape anomalies.

\section{Acknowledgements}
This work was supported by the CNRS challenge NEEDS and the associated NACRE project, as well as the CHANDA FP7/EURATOM project (Contract No. 605203) and the Spanish Ministerio
de Economia y Competitividad under grants FPA2014-52823-C2-1, FPA2017-83946-C2-1 and SEV-2014-0398.
We acknowledge the support of STFC (UK) council grant ST/P005314/1. The authors thank T. Yoshida, T. Tachibana, S. Okumura, and S. Chiba for providing the dataset from their most recent gross theory calculations and the Nuclear Data Section of the International Atomic Energy Agency for fostering the advances of nuclear data and their critical assessment.

\end{document}